\documentclass[review]{elsarticle}

\usepackage{lineno,hyperref}

\usepackage{amsmath,amssymb,amsthm}

\usepackage{subcaption}
\newlength\savedwidth

\newtheorem{theorem}{Theorem}
\newtheorem{lemma}[theorem]{Lemma}

\modulolinenumbers[5]

\journal{Journal of \LaTeX\ Templates}

\usepackage{color}
\usepackage{soul}

\begin{document}

\begin{frontmatter}

\title{Tuning the Clustering Coefficient of Generalized Circulant Networks}



\author[robaddress]{Robert E. Kooij} 
\author[sutdaddress]{Nikolaj Horsevad S{\o}rensen}
\author[uoaddress]{Roland Bouffanais\corref{mycorrespondingauthor}}
\cortext[mycorrespondingauthor]{Corresponding author}
\ead{rbouffan@uottawa.ca}

\address[robaddress]{Delft University of Technology, Delft, The Netherlands}
\address[sutdaddress]{Singapore University of Technology and Design, Singapore, Singapore}
\address[uoaddress]{Faculty of Engineering, University of Ottawa, Ottawa, Canada}

\begin{abstract}
Apart from the role the clustering coefficient plays in the definition of the small-world phenomena, it also has great relevance for practical problems involving networked dynamical systems. To study the impact of the clustering coefficient on dynamical processes taking place on networks, some authors have focused on the construction of graphs with tunable clustering coefficients. These constructions are usually realized through a stochastic process, either by growing a network through the preferential attachment procedure, or by applying a random rewiring process. In contrast, we consider here several families of static graphs whose clustering coefficients can be determined explicitly. The basis for these families is formed by the $k$-regular graphs on $N$ nodes, that belong to the family of so-called circulant graphs denoted by $C_{N,k}$. We show that the expression for the clustering coefficient of $C_{N,k}$ reported in literature, only holds for sufficiently large $N$. Next, we consider three generalizations of the circulant graphs, either by adding some pendant links to $C_{N,k}$, or by connecting, in two different ways, an additional node to some nodes of $C_{N,k}$. For all three generalizations, we derive explicit expressions for the clustering coefficient. Finally, we construct a family of pairs of generalized circulant graphs, with the same number of nodes and links, but with different clustering coefficients. 
\end{abstract}

\begin{keyword}
Circulant Graphs \sep Clustered Networks  \sep Consensus Dynamics \sep Complex Networks 
\end{keyword}

\end{frontmatter}


\section{Introduction}

To characterize key features of networks across a wide range of applications, network science has developed a wide range of metrics and mathematical indices~\cite{barrat2008dynamical}. Among the most commonly used metrics are the average shortest path length, degree distribution and clustering coefficient. The latter index was introduced by Watts \& Strogatz in their seminal paper introducing the concept of ``small-world" networks~\cite{watts1998collective}. Specifically, nodes in a network are characterized by the structure of their local neighborhood or connectivity. The concept of clustering is associated with the tendency of some nodes to form cliques as observed in many real-world networks. This index accounts for the ratio between the actual number of relations among the neighbors of a node and the maximum possible number of such relations. The clustering coefficient is defined as the average of this ratio, when averaged over all nodes of the network.

Although the clustering coefficient is generally high in many small-world social and technological networks, it would be wrong to directly associate the small-world feature with high levels of clustering. For instance, random networks yield short path lengths---i.e. they exhibit the small-world feature---but fail to achieve high levels of clustering due to the random interconnectivity of nodes, which hinders the process of clique formation~\cite{barrat2008dynamical}. The empirical observation of very high levels of clustering in many real-world networks has therefore been a source of inquiry and to deal with this conceptual challenge, many network models with tunable clustering coefficient have been proposed~\cite{holme2002growing,sekunda2016}. For instance, varying clustering coefficients are encountered in gene expression networks~\cite{kalna2007clustering}, and also in node spreading influence~\cite{wang2017identifying}. The spreading of complex contagious processes---e.g. social cooperation, protest events, the spread of consumer goods and technological innovations, the diffusion of health care preventive measures, the growth of violent-crime epidemics, the spread of institutional norms, the adoption of cultural practices, etc.---has been clearly tied to particular levels of clustering in social and co-presence networks~\cite{centola2007cascade}. 

To study the impact of the clustering coefficient on dynamical processes taking place on any network, many authors have focused on the construction of graphs with tunable clustering coefficient, see for instance~\cite{holme2002growing,sekunda2016}. Typically, the process to reach a desired value for the clustering coefficient is realized through a stochastic growth process, e.g. following the preferential attachment procedure~\cite{newman2001clustering} or alternatively through the rewiring of regular network lattices~{\cite{watts1998collective}}. Due to the stochastic nature of these network models most of the results concerning the network properties are only valid for large networks. 

In what follows, we give a summary of the state of the art associated with the construction of graphs with tunable clustering coefficient. In~{\cite{MU201955}}, the authors explore a ``local" rewiring strategy, which allows for an efficient way to adjust the clustering coefficient, and steer it towards a desired value. However, this method depends on randomness in the choice of link pairs that are rewired and thus it is not known a priori if a certain target value for the clustering coefficient can effectively be attained. Kr\"uger et al.~{\cite{Kr_ger_2015}} use an approach of random ``triangulations" of lattices and determine, amongst others, the clustering coefficient. However, the stochastic nature of the process that determines these so-called triangulations of the lattice imposes that the obtained results for the clustering coefficient are essentially (ensemble) average properties, and thus are only valid for large networks. A simple family of growing small-world networks is proposed by Shang~{\cite{articleShang}}. The constructed graphs depends on growing the network by closing triangles, followed by a random process of link removal. By tuning the link removal, networks with different clustering coefficients but similar degree distribution are obtained. However, in this family of networks the clustering coefficient can only be determined explicitly for the extreme cases of the model, namely the path graph and the so-called Faray graphs. In~{\cite{yang2014}}, Yang et al. introduced a model that generates power-law distributions of degree and a tunable clustering coefficient. However, the construction process is based upon preferential attachment and hence stochastic in nature. In addition, the constructed networks are necessarily weighted. A variation of the Configuration Model (CM), which also takes the formation of triangles into account is proposed in~{\cite{Serrano}}. Thanks to this approach, networks can be generated with a given degree distribution and a preassigned degree-dependent clustering coefficient. Unfortunately, for small-size networks multi-links and self-loops can occur, which often have to be discarded for some practical applications. Furthermore, the CM uses randomly selected stubs to generate links, so there is no control over the final network topology. The Block Two-Level Erd\H{o}s-R\'{e}nyi model (BTER) is introduced in~{\cite{BTER}}. The BTER model is derived to mimic real-world networks exhibiting community structure. This model also is based upon stochasticity, and again, the results are fundamentally asymptotic for large networks. In addition Ref.~\cite{BTER} focuses on the transitivity index and not the clustering coefficient. In summary, the literature offers numerous models and family of networks with some forms of tuning or adjustment of the level of clustering. However, the vast majority of the available network-generating models are intrinsically stochastic in nature, and often assume large networks, with the control over the clustering coefficient and other network properties to be valid only over ensemble averaging.

In contrast, in this paper we consider a family of deterministic graphs whose clustering coefficient can be determined explicitly and exactly. This study is partially motivated by a number of practical applications in the field of networked robotics. Recently, Mateo et al.~{\cite{mateo2019}} suggested to study the impact of the clustering coefficient on the responsiveness of multi-agent systems performing distributed consensus. Their results were backed up by experiments using swarm robotic networks, albeit these systems are small in size~{\cite{mateo2019,zoss2018,kit2019}}. Thus in order to conduct these experiments, it is important to have explicit expressions for the clustering coefficient, such that its impact on the dynamic processes considered can be easily determined. Since the network size is rather small---with 10 to 50 robots---the asymptotic properties of stochastic models cannot reliably be used.

The basis for our model is formed by $k$-regular graphs on $N$ nodes, that belong to the family of so-called circulant graphs. In this family of graphs, which we will denote by $C_{N,k}$, the $N$ nodes form a cycle graph $C_N$. In addition, every node also connects to all nodes at distance at most $k/2$ on the cycle, where it is assumed that $k$ is even. This graph forms the basis for the construction of the Watts \& Strogatz small-world graphs~\cite{watts1998collective} and the clustering coefficient for $C_{N,k}$ is given by Newman in~\cite{newman2003structure}. However, we will show that the formula given by~\cite{newman2003structure} only holds if $N$ is sufficiently large.

In addition, if one assumes that both ends of each edge are rewired with probability $p$, and by allowing both double and self-loops, then the clustering of the resulting small-world graph can also be determined~\cite{barrat2000properties}. The corresponding formula reported in Ref.~\cite{barrat2000properties} is the product of the clustering coefficient for $C_{N,k}$ and the factor $(1-p)^3$. It is worth noting that this formula is also incomplete in the sense that it presumes $N$ to be sufficiently large.
Although both Refs.~\cite{holme2002growing} and \cite{barrat2000properties} provide a way to construct graphs with a given clustering coefficient, there is no control over the resulting graph given the fact that their methods are based upon a stochastic process. Moreover, the construction suggested by Ref.~\cite{barrat2000properties} has a non-negligible probability to lead to graphs with multiple links and self-loops.

In this paper, we determine the complete analytical expressions for the clustering coefficient of $C_{N,k}$ for any value of $N$. We also consider three generalizations of the circulant graphs $C_{N,k}$. For the first generalization, we add a number of pendant links to $C_{N,k}$. For the second generalization, we add a single node to $C_{N,k}$, such that it connects to $m$ nodes of $C_{N,k}$, which are adjacent on $C_N$. Finally, the third generalization comprises of $C_{N,k}$ with one single node added which connects to $m$ nodes on $C_{N,k}$, which are exactly $1+\frac{k}{2}$ hops away on the cycle $C_N$. For all three generalizations we derive explicit expressions for the clustering coefficient. The present article is partially motivated by the fact that a generalized circulant graph of Type I has the highest collective response at moderate frequency values, in a leader-follower system consisting of 11 agents, see Fig. 2 in ~\cite{mateo2019}. 
Note that in~{\cite{circulants}}, the authors propose a generalization of a class of recursive circulant graphs. However, it is worth pointing out that these generalized recursive circulant graphs belong to the set of circulant graphs, and hence are all vertex-symmetric. On the other hand, our generalized circulant graphs, do not belong the the class of circulant graphs, as they are not vertex-symmetric.

Lastly, we demonstrate how this family of graphs can be used to create a set of particular network topologies of the exact same size and number of edges, yet exhibiting vastly different levels of clustering. We demonstrate the impact of changing the network topology for a particular network dynamic, namely the leader-follower consensus, which shows how networked systems with a small number of nodes $N$ are also strongly influenced by the chosen topology, and thus having a family of networks with explicitly known graph-metrics is of great use. 

\section{Clustering coefficient for the circulant graphs $C_{N,k}$}

For the computation of the clustering coefficient, two cases have to be considered. This becomes clear upon introducing some notations. Let the nodes of $C_{N,k}$ be labeled clockwise from~1~to~$N$. Without loss of generality, let us focus on the clustering coefficient of node~1. It is obvious that node 1 is connected to the set of nodes $S_1 = \{2,3,\cdots,k/2+1\}$ and to the node set $S_2 =\{N-k/2+1,\cdots,N-1,N\}$. Now, if the distance between the last node in $S_1$, i.e. node $k/2+1$ and the first node in $S_2$, i.e. $N-k/2+1$, is more than $k/2$ on the cycle $C_N$, then node $k/2+1$ does not connect to any node in $S_2$. 
This condition translates to
\begin{equation}\label{cond1}
N > \frac{3k}{2}.    
\end{equation}

We will now determine the clustering coefficient for node 1, under condition~\eqref{cond1}. Obviously, node 1 has a degree $k$ by construction. Therefore, the maximum number of connections between its neighbors is ${k}\choose{2}$. We now determine the number of connections between the neighbors of node 1, that are missing. Starting with node  $k/2+1$, we have seen that $k/2$ links are missing. Next, node $k/2$, is missing $k/2-1$ links. Repeating this argument until node 2, we arrive at a total of $k/2+1 + k/2 +\cdots + 1 = \frac{k}{4}(\frac{k}{2}+1)$ missing links. In conclusion, the clustering $C_1$ of node 1 satisfies $C_1 = \frac{ {{k} \choose {2}} -\frac{k}{4}(\frac{k}{2}+1)}
{{{k} \choose {2}}}$. Given the symmetry of $C_{N,k}$, all nodes of this graph have an identical clustering coefficient. Simplifying the expression for $C_1$ we arrive at the following result.

\begin{lemma}\label{T1}
Consider the circulant graph $C_{N,k}$, satisfying the condition $N > \frac{3k}{2}$. Then the clustering coefficient $C$ of $C_{N,k}$ satisfies \begin{equation}\label{C1}
C=\frac{3(k-2)}{4(k-1)}.
\end{equation}
\end{lemma}

\begin{figure}[htbp]
\centering
\includegraphics[]{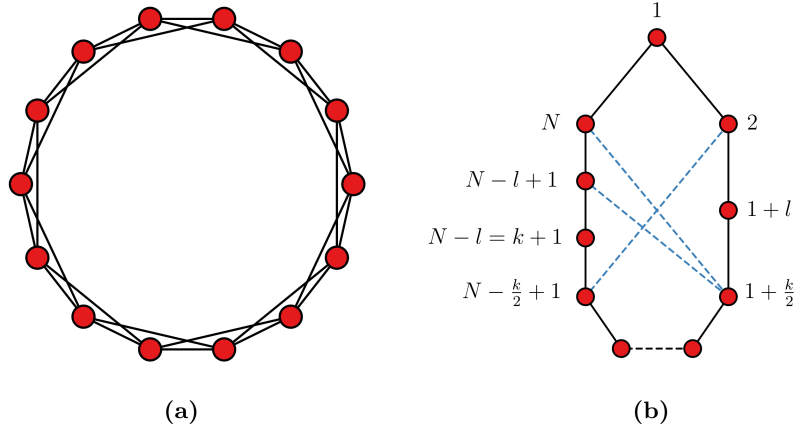}
\caption{(a) Circulant graph $C_{14,4}$ satisfying the condition $N > \frac{3k}{2}$.; (b) Relative positions of nodes $1+k=N-l$ and $N+1-k/2$.}\label{fig:1}
\end{figure}

Note that Eq.~\eqref{C1} is given in~\cite{newman2003structure}, albeit without mention of condition~\eqref{cond1}. If $k$ becomes large, then $C$ approaches 3/4. This is why small-world networks constructed be rewiring a small percentage of the links in $C_{N,k}$, have a clustering coefficient close to 3/4 for large $k$ (see also Ref.~\cite{watts1998collective}). Figure~\ref{fig:1}a depicts an example of a circulant graph satisfying the condition of Lemma \ref{T1}.

We now turn our attention to the remaining case:
\begin{equation}\label{cond2}
N \leq \frac{3k}{2}.    
\end{equation}%
Again, without loss of generality, we analyze the clustering coefficient of node 1. Starting from the last node in the set $S_1$, i.e. node $1+k/2$, and moving clockwise $k/2$ hops on the cycle $C_N$, we arrive at node $1+k$. Under condition~\eqref{cond2}, this node belongs to the set $S_2$ (see Fig.~\ref{fig:1}b).

For convenience sake, we introduce $l=N-k-1$ with $l \geq 0$ since $C_{N,k}$ contains at least $k+1$ nodes. Obviously, node $1+k$ then corresponds to node $N-l$. As a result, moving clockwise on $C_N$ from  $1+k/2$, we cannot reach nodes $N-l+1,\cdots,N$ in set $S_2$. Therefore, the number of nodes in $S_2$ that cannot be reached by node $1+k/2$ is $l$. The same holds for the nodes $1+l,\cdots,k/2$. Next, node $l$ in set $S_1$ does not reach $l-1$ nodes in $S_2$. Continuing this argument, we arrive at node $2$ in set $S_1$, that only does not reach one node in $S_2$, namely node $N-k/2+1$. In summary, the total number of missing connections between sets $S_1$ and $S_2$ is given by $l\left(\frac{k}{2}-l+1\right)+l-1+l-2+\cdots+1=\frac{kl}{2}-\frac{l(l-1)}{2}$. Again, using symmetry arguments for $C_{N,k}$, we arrive at the following result.

\begin{lemma}\label{T2}
Consider the circulant graph $C_{N,k}$, satisfying the condition $N \leq \frac{3k}{2}$. Then the clustering coefficient $C$ satisfies \begin{equation}\label{C2}
C = \frac{{{k} \choose {2}} -\frac{kl}{2}+\frac{l(l-1)}{2}}{{{k} \choose {2}}},
\end{equation}
where $l=N-k-1$.
\end{lemma}

\begin{figure}[htbp]
\centering
\includegraphics[width=0.4\textwidth]{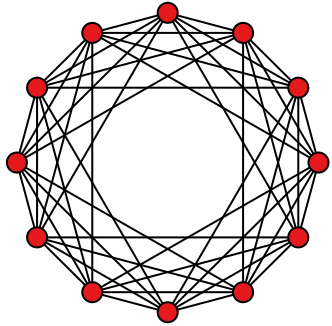}
\caption{Circulant graph $C_{12,8}$ satisfying the condition $N \leq \frac{3k}{2}$.}\label{fig:2}
\end{figure}

Note that for the particular case $l=0$, i.e. $N=k+1$, the circulant $C_{N,k}$ is a complete graph, which therefor has the clustering coefficient $C=1$. Figure~\ref{fig:2} depicts an example of a circulant graph satisfying the condition of Lemma \ref{T2}.

Combining Lemmas \ref{T1} and \ref{T2}, we obtain the following general result.
\begin{theorem}\label{T3}
Consider the circulant graph $C_{N,k}$, and define $r=\min \{N-k-1,\frac{k}{2} \}$. Then the clustering coefficient $C$ satisfies \begin{equation}\label{C3}
C = \frac{{{k} \choose {2}} -\frac{kr}{2}+\frac{r(r-1)}{2}}{{{k} \choose {2}}}.
\end{equation}
\end{theorem}

\textbf{Proof.}
First, assume that condition \eqref{cond1} holds. Then, it follows $N-k-1 > k/2$, hence $r=k/2$. An elementary calculation reveals that substitution of $r=k/2$ into Eq.~\eqref{C3} gives Eq.~\eqref{C1}. Finally, if condition (\ref{cond2}) holds, it follows that $N-k-1 \leq k/2$, hence $r=N-k-1=l$. Now Eq.~\eqref{C3} corresponds to Eq.~\eqref{C2}. This finishes the proof. \qed  
\linebreak

\begin{figure}[htbp]
\centering
\includegraphics[width=0.8\textwidth]{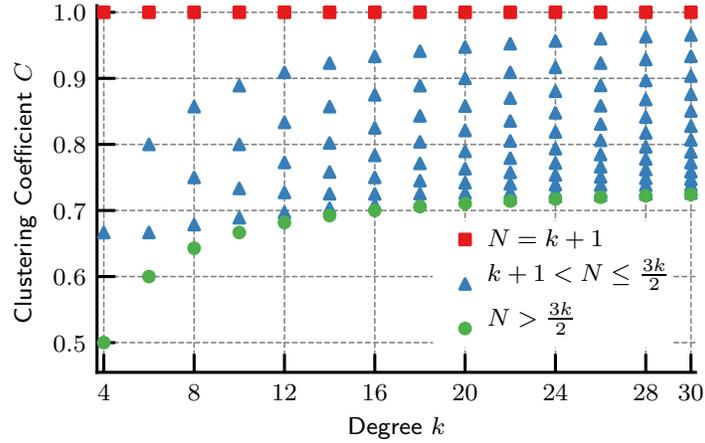}
\caption{All possible values of clustering coefficient $C$ for $C_{N,k}$ with $k \leq 30$.}\label{fig:3}
\end{figure}

The above general expression~\eqref{C3} for the the clustering coefficient of circulant graphs allows us to obtain the inequalities $\frac{3(k-2)}{4(k-1)} \leq C \leq 1 $ for any given $k$ and $C_{N,k}$. Furthermore, $C$ attains $k/2 - 1$ values between the given bounds. Figure~\ref{fig:3} depicts this for values of $k$ up to 30.

\section{Clustering coefficient for the generalized circulant graphs $C_{N,k,m,s}$ of Type I}

\begin{figure}[htbp]
\centering
\includegraphics[]{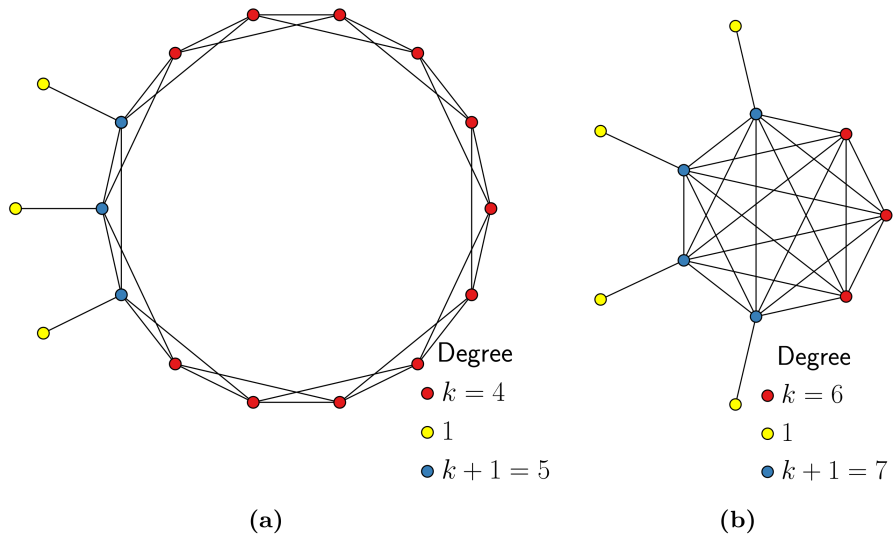}
\caption{(a) The graph $C_{14,4,3}$ from the Type I subfamily of circulant graphs.; (b) The generalized circulant graph $C_{7,6,4}$.}\label{fig:4}
\end{figure}

We have seen in the previous section that for the circulant graph $C_{N,k}$, the clustering coefficient is at least equal to $\frac{3(k-2)}{4(k-1)}$. We now consider a modification to the family of circulant graphs in order to bring down the value of the cluster coefficient. To achieve this, we add $m$ ``leaves" to $C_{N,k}$, i.e. we add $m$ degree-one nodes to the graph---with $m \leq N$---and attach each node to a separate node on $C_{N,k}$. We denote the obtained family of graphs by $C_{N,k,m}$. As an example, Fig.~\ref{fig:4}a depicts $C_{14,4,3}$.

Mateo et al.~\cite{mateo2019} studied topologies that optimize the collective response of multi-agent systems subjected to a dynamic linear leader-follower consensus protocol. Using simulated annealing optimization for the frequency response of a system of $N+1=11$ networked agents---one leader and 10 followers, they found that for moderate values of the frequency $\omega$ of the variations of the leading agent, the generalized circulant graph $C_{7,6,4}$ (see Fig.~\ref{fig:4}b and Fig. 2 in~\cite{mateo2019}) has the highest collective response at a frequency of $\omega=0.4$~\cite{mateo2019}.

\begin{theorem}\label{T4}
Consider the graph $C_{N,k,m}$ and define $r=\min \{N-k-1,\frac{k}{2} \}$ and $A={{k} \choose {2}} -\frac{kr}{2}+\frac{r(r-1)}{2}$. Then the clustering coefficient $C$ satisfies \begin{equation}\label{C4}
C = \frac{2A(Nk+N-2m)}{(N+m)k(k^2-1)}.
\end{equation}
\end{theorem}

\textbf{Proof.} The $m$ nodes attached to $C_{N,k}$ obviously have a clustering of $0$ since they are of degree one. The $N-m$ nodes that are not connected to the $m$ nodes, have the same clustering as in the case of $C_{N,k}$. Using $r=\min \{N-k-1,\frac{k}{2} \}$ and $A={{{k} \choose {2}}} -\frac{kr}{2}+\frac{r(r-1)}{2}$, according to Theorem \ref{T3}, their clustering can be written as $\frac{A}{{{k} \choose {2}}}$. Finally, we consider the $m$ nodes on $C_{N,k}$ that are connected to one of the $m$ pendant nodes (yellow nodes in Fig.~\ref{fig:4}). These nodes have $k+1$ neighbors but the number of connections between these neighbors is the same as for the $C_{N,k}$ case. Therefore, the clustering for these $m$ nodes satisfies $\frac{A}{{{k+1} \choose {2}}}$. Summing over all nodes we obtain
$$C=\frac{(N-m)\frac{A}{{k \choose 2}}+m\frac{A}{{{k+1} \choose 2}}}{N+m}.$$ Factorizing the latter expression leads to Eq.~\eqref{C4} and finishes the proof. $ \qed $
\linebreak

\begin{figure}[htbp]
\centering
\includegraphics[width=0.8\textwidth]{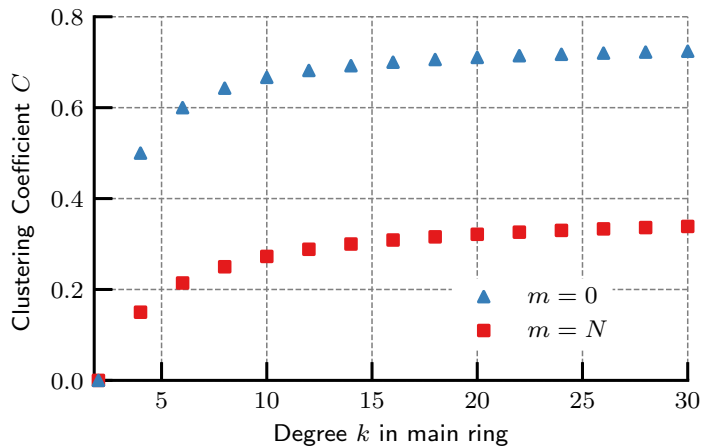}
\caption{Envelope for the values of the clustering coefficient of the generalized circulant graphs $C_{N,k,m}$ with $N >\frac{3k}{2}$.}\label{fig:5}
\end{figure}

Note that for $m=0$, we retrieve the result stated in Theorem \ref{T3}. For any given $N$ and $k$, upon increasing $m$, the clustering coefficient does become smaller. The smallest value is obtained for $m=N$. It is easy to verify that for $N > \frac{3k}{2}$ the clustering coefficient of $C_{N,k,N}$ satisfies
\begin{equation}\label{C5}
C = \frac{3(k-2)}{8(k+1)}.
\end{equation}
Therefore, under the condition $N > \frac{3k}{2}$, the values $\frac{3(k-2)}{8(k+1)}$ and $\frac{3(k-2)}{4(k-1)}$ form an envelope for the clustering coefficient for $C_{N,k,m}$ (see Fig.~\ref{fig:5}). For each $k$, by varying $m$ from $1$ to $N-1$, we can obtain $N-1$ values for $C$ within the envelope depicted in Fig.~\ref{fig:5}. 

It is clear from the above result, that a lower bound for the clustering coefficient for $C_{N,k,m}$ is given by Eq.~\eqref{C5}. If one seeks to generate clustering coefficients even lower than this value, one has to further modify the topology of the circulant graphs. Intuitively, this can be achieved by adding more pendant nodes. To do this in a structured way, we propose the following generalization for the $C_{N,k,m}$  construction. 

\begin{figure}[htbp]
\centering
\includegraphics[width=0.575\textwidth]{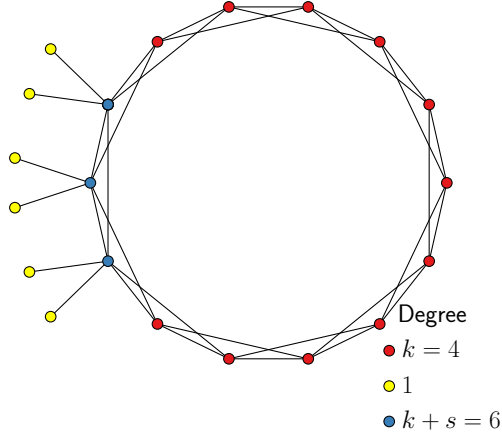}
\caption{Generalized circulant graph of Type I $C_{14,4,3,2}$.}\label{fig:6}
\end{figure}

We start with the circulant graphs family $C_{N,k}$. To $m$ nodes in $C_{N,k}$, we add $s$ pendant nodes with $s \geq 1$ and $m \leq N$. We denote the obtained family of graphs $C_{N,k,m,s}$ and refer to is as a generalized circulant graph of Type I. As an example, Fig.~\ref{fig:6} depicts $C_{14,4,3,2}$.

\begin{theorem}\label{T5}
Consider the generalized circulant graph $C_{N,k,m,s}$ of Type I and define $r=\min \{N-k-1,\frac{k}{2} \}$ and $A={{k} \choose {2}} -\frac{kr}{2}+\frac{r(r-1)}{2}$. Then the clustering coefficient $C$ satisfies \begin{equation}\label{C6}
C = \frac{2A}{N+ms}\left( \frac{N-m}{k(k-1)}+\frac{m}{(k+s)(k+s-1)} \right).
\end{equation}
\end{theorem}

\begin{figure}[htbp]
\centering
\includegraphics[width=0.8\textwidth]{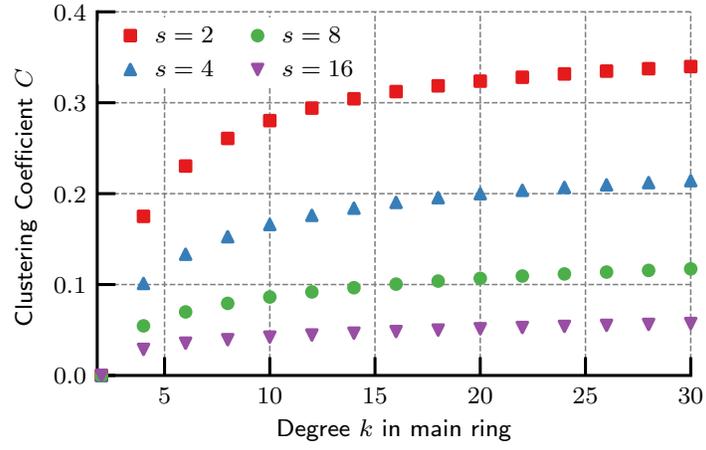}
\caption{Clustering coefficient for generalized circulant graph of Type I, $C_{  N,k,m,s}$, with $N > \frac{3k}{2}$.}\label{fig:7}
\end{figure}

The proof follows from arguments very similar to those presented in the proof of Theorem \ref{T4} and is therefore omitted.
\newline    
\newline
For any given $N, k$ and $s$,  the smallest value for the clustering coefficient is obtained for $m=N$. It is straightforward to show that for $N > \frac{3k}{2}$ the clustering coefficient of $C_{N,k,N,s}$ satisfies
\begin{equation}\label{C7}
C = \frac{3k(k-2)}{4(s+1)(k+s)(k+s-1)}.
\end{equation}
Figure~\ref{fig:7} shows values of the clustering coefficient for $C_{N,k,N,s}$ as a function of $k$, with $N > \frac{3k}{2}$ and for several values of $s$.

\section{Clustering coefficient for the generalized circulant graphs $D_{N,k,m}$ of Type II}

\begin{figure}[htbp]
\captionsetup[subfigure]{justification=centering}
\centering
\includegraphics[]{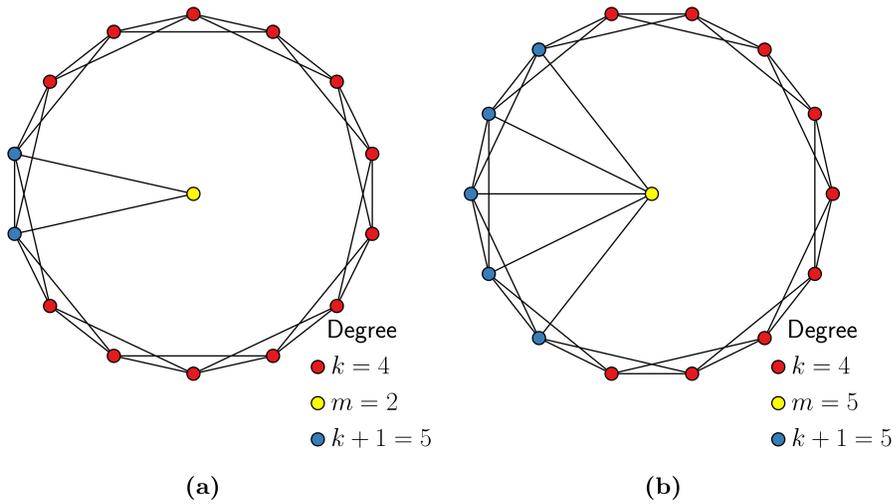}
\caption{Generalized circulant graphs of Type II: $D_{N,k,m}$. Particular cases of $D_{14,4,2}$ (a) and $D_{14,4,5}$ (b). }\label{fig:8}
\end{figure}

In this section, we consider another type of alteration of the main topology of circulant graphs $C_{N,k}$, but this time by adding one single node, which connects to $m$ nodes on $C_{N,k}$, with $1 \leq m \leq N$. It is assumed that the $m$ nodes on $C_{N,k}$ are adjacent on $C_N$. We call the obtained family of graphs, a generalized circulant graph of Type II and denote it by $D_{N,k,m}$. Two examples are depicted in Fig.~\ref{fig:8}: $D_{14,4,2}$ and $D_{14,4,5}$.

To determine the clustering coefficient of $D_{N,k,m}$, we have yet again to consider two cases. For the first case, all $m$ neighbors of the central node are connected. This case occurs for $m \leq \frac{k}{2} + 1$. For instance, the particular case $D_{14,4,2}$ shown in Fig.~\ref{fig:8}a fits this case. For this case, if $m>1$, the clustering of the central node is 1. Note that for $m=1$, $D_{N,k,m}$ is equal to $C_{N,k,1}$, so the clustering coefficient for this case follows from the results obtained in the previous section.

\begin{lemma}\label{T6}
Consider the generalized circulant graph $D_{N,k,m}$ of Type II, with $1 < m \leq \frac{k}{2} + 1$. Define $r=\min \{N-k-1,\frac{k}{2} \}$ and $A={{k} \choose {2}} -\frac{kr}{2}+\frac{r(r-1)}{2}$. Then the clustering coefficient $C$ satisfies 
\begin{equation}\label{C10}
C=\frac{(N-m)\frac{A}{{k \choose 2}}+m\frac{(A+m-1)}{{{k+1} \choose 2}}+1}{N+1}.
\end{equation}
\end{lemma}

\textbf{Proof.} The $N-m$ nodes on $C_{N,k}$ that are not connected to the central node, have the same clustering as in $C_{N,k}$, namely $\frac{A}{{{k} \choose {2}}}$. Denote the set of $m$ nodes on $D_{N,k,m}$ which are connected to the central node by $I$. A node $i$ in $I$ has $k+1$ neighbors. Because $m \leq \frac{k}{2} + 1$, among those neighbors there are $A$ connections which are part of $C_{N,k}$. There are $m-1$ additional connections between the neighbors of node $i$, namely from the central node to the nodes in $I$. Hence each node $i$ in $I$ has clustering 
$C_i=\frac{A+m-1}{{{k+1} \choose 2}}$. 
As mentioned before, the central node has clustering $1$, for $m \leq \frac{k}{2} + 1$. Summing over all nodes, we obtain Eq.~\eqref{C10}. This completes the proof. $ \qed $
\linebreak

\begin{lemma}\label{T7}
Consider the generalized circulant graph $D_{N,k,m}$ of Type II, with $m > \frac{k}{2} + 1$. Define $r=\min \{N-k-1,\frac{k}{2} \}$ and $A={{k} \choose {2}} -\frac{kr}{2}+\frac{r(r-1)}{2}$. Then the clustering coefficient $C$ satisfies 
\begin{equation}\label{C7bis}
C=\frac{(N-m)\frac{A}{{k \choose 2}}+\frac{mA+\frac{k}{4}(3k-2)+(m-k)k}{{{k+1} \choose 2}}+\frac{{{m} \choose 2}-\frac{(m-\frac{k}{2})(m-\frac{k}{2}-1)}{2}}{{{m} \choose 2}}}{N+1}.
\end{equation}

\end{lemma}

\begin{figure}[htbp]
\centering
\includegraphics[width=0.5\textwidth]{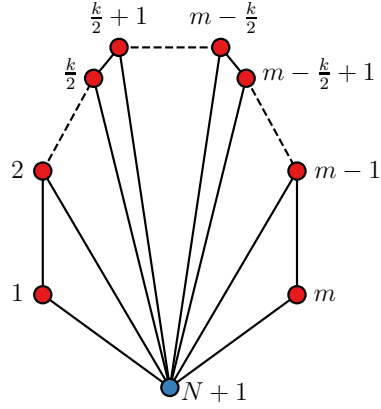}
\caption{Relative positions of nodes for the case $m \geq k + 1$.}\label{fig:9}
\end{figure}

\textbf{Proof.} Following the same train of thought as in the proof of Lemma~\ref{T6}, the $N-m$ nodes on $C_{N,k}$ that are not connected to the central node, have a clustering $\frac{A}{{{k} \choose {2}}}$. When considering the set of nodes $I$ on $D_{N,k,m}$ which are connected to the central node, it is convenient to consider two cases. We start with the case $m \geq k+1$ (see Fig.~\ref{fig:9}).

Denote the nodes with index up to $\frac{k}{2}$ as $I_1$ and those with index $\frac{k}{2}+1$ up to index $m-\frac{k}{2}$ as $I_2$. The remainder of the nodes in $I$ are denoted as $I_3$. We will now count the number of connections between the neighbors of all nodes in $I$, in addition to the $A$ connections which are part of $C_{N,k}$. Node $1$ in $I_1$ induces $\frac{k}{2}$ additional connections. Node $2$ induces $\frac{k}{2}+1$ additional connections. It can be deduced that the total number of additional connections due to $I_1$ is given by 
$\frac{k}{2} + \frac{k}{2} + 1 + \cdots + \frac{k}{2} + \frac{k}{2} - 1 = \frac{3k}{8}(3k-2)$.
Every node in $I_2$ induces $k$ additional connections, hence $I_2$ contributes $(m-k)k$ additional connections. Finally, the contribution of $I_3$ is similar to the contribution of $I_1$, by symmetry. Therefore, for the case $m \geq k + 1$, the contribution to the clustering of all nodes in $I$ amounts to 
\begin{equation}\label{C12}
C_I=\frac{mA+\frac{k}{4}(3k-2)+(m-k)k}{{{k+1} \choose 2}}.
\end{equation}
For the case $m < k + 1$, the set of nodes $I$, can be split into two sets: the aforementioned set $I_1$ and the remaining nodes in $I$. With an argumentation similar to the one used in the case $m \geq k+1$, it can be shown that the contribution of the nodes in $I$ to the clustering, is also given by Eq. (\ref{C12}). We leave the details as an exercise to the reader.  
\newline
Finally, we look at the clustering of the central node. We count the number of connections between its $m$ neighbors, which are missing. Starting with node 1, we miss $m - \frac{k}{2} -1$ connections. For node 2, we miss $m - \frac{k}{2} -2$ connections. and so on. Therefore the total number of missing connections satisfies $(m - \frac{k}{2} -1) + (m - \frac{k}{2} -2) + \cdots + 1 = \frac{(m-\frac{k}{2})(m-\frac{k}{2}-1)}{2}$.
Therefore, the clustering of the central node is given by 
$\frac{{{m} \choose 2}-\frac{(m-\frac{k}{2})(m-\frac{k}{2}-1)}{2}}{{{m} \choose 2}}$.
Summing over all nodes we obtain Eq.~\eqref{C7bis}. $\qed$
\linebreak

\begin{theorem}\label{T8}
Consider the generalized circulant graph $D_{N,k,m}$ of Type II, with $m > 1$. Define $r=\min \{N-k-1,\frac{k}{2} \}, A={{k} \choose {2}} -\frac{kr}{2}+\frac{r(r-1)}{2}$ and let $\sigma=min\{m-1,\frac{k}{2}\}$. Then the clustering coefficient $C$ satisfies 
\begin{equation}\label{C8}
C=\frac{(N-m)\frac{A}{{k \choose 2}}+\frac{mA+\sigma(3\sigma -1)+2(m-2\sigma)\sigma}{{{k+1} \choose 2}}+\frac{{{m} \choose 2}-\frac{(m-\frac{k}{2})(m-1-\sigma)}{2}}{{{m} \choose 2}}}{N+1}
\end{equation}
\end{theorem}
 
\textbf{Proof.}
First, assume that the condition $m \leq \frac{k}{2}+1$ holds. Then, $\sigma=m-1$. Substitution of this value in Eq.~\eqref{C8} gives Eq.~\eqref{C10}. Finally, if $m > \frac{k}{2}+1$, then, $\sigma=\frac{k}{2}$. Plugging this value into Eq.~\eqref{C8} leads to Eq.~\eqref{C7bis}. This finishes the proof. $\qed$
\linebreak

\begin{figure}[htbp]
\centering          
\includegraphics[width=0.8\textwidth]{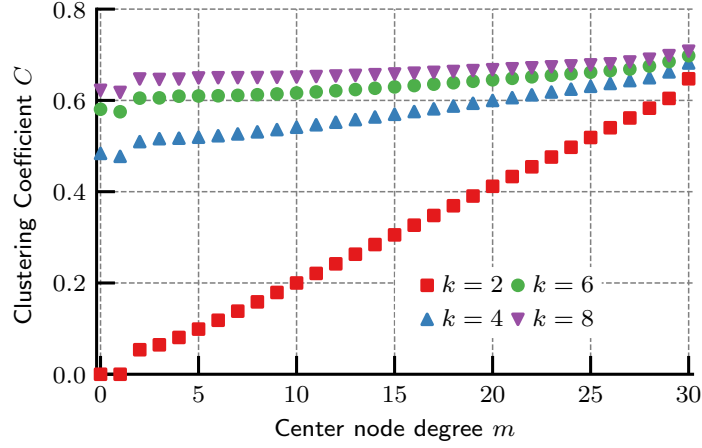}
\caption{Values of the clustering coefficient $C$ for the generalized circulant graph of Type II $D_{30,k,m}$ when varying the degree of the central node.}\label{fig:10}
\end{figure}

Figure~\ref{fig:10} depicts the clustering coefficient of $D_{30,k,m}$, with $m \in \{0,1,\dots,30\}$. 
We observe the clustering coefficient drops for $m=1$, example: $C(D_{30,4,1}) < C(C_{30,4})=\frac{1}{2}$, while for $m > 1$ it holds that $C(D_{30,4,m} )> C(C_{30,4})$.

\section{Clustering coefficient for the generalized circulant graphs $E_{N,k,m}$ of Type III}

In the previous section, we have generalized the circulant graph $C_{N,k}$ by adding one central node that connects to $m$ adjacent nodes on the circulant graph. We have seen an example of the resulting circulant graph $D_{N,k,m}$ of Type II, with a higher clustering coefficient than $C_{N,k}$, for $m>1$. In this section, we introduce another generalization of $C_{N,k}$, again with one central node, but this time leading to smaller values for the clustering coefficient.

\begin{figure}[htbp]
\centering
\includegraphics[]{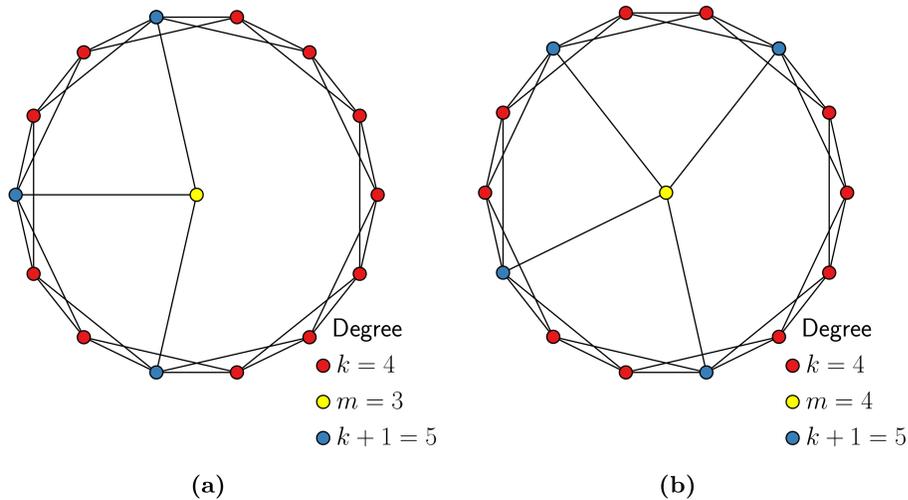}
\caption{Generalized circulant graphs of Type III $E_{N,k,m}$. The particular cases are $E_{14,4,3}$ (a) and $E_{14,4,4 }$ (b).}\label{fig:11}
\end{figure}

We consider adding one central node to the family of circulant graphs $C_{N,k}$. The node will connect to $m$ nodes on $C_{N,k}$, which are exactly $1+\frac{k}{2}$ hops away on the cycle $C_N$. Therefore, $m$ has to satisfy $m \leq     \left\lfloor\frac{N}{1+\frac{k}{2}}\right\rfloor$. We call the obtained graph, a generalized circulant graph of Type III and denote it by $E_{N,k,m}$. Two examples are shown in Fig.~\ref{fig:11}: $E_{14,4,3}$ and $E_{14,4,4 }$.

\begin{theorem}\label{T9}
Consider the generalized circulant graph $E_{N,k,m}$ of Type III and define $r=\min \{N-k-1,\frac{k}{2} \}$ and $A={{k} \choose {2}} -\frac{kr}{2}+\frac{r(r-1)}{2}$. Then the clustering coefficient $C$ satisfies 
\begin{equation}\label{C14}
C=\frac{(N-m)\frac{A}{{k \choose 2}}+\frac{mA}{{{k+1} \choose 2}}}{N+1}.
\end{equation}
\end{theorem}

The proof of Theorem \ref{T9} is based on similar arguments as the proof of the previous theorems and is therefore omitted.
\newline
It is easy to see from Eq.~\eqref{C14} that $C$ is a linearly decreasing function of $m$. Similarly, subtracting Eq.~\eqref{C14} from Eq.~\eqref{C3}  , we can show that the clustering coefficient of $E_{N,k,m}$ is always smaller than that of $C_{N,k}$.

\section{Discussion}

In Ref.~\cite{mateo2019}, Mateo et al. investigated the influence of the network topology on collective response. Specifically, they considered an archetypal model of distributed decision-making---the leader-follower linear consensus---and studied the collective capacity of the system to follow a dynamic driving signal (the ``leader", an external perturbation locally disturbing the collective dynamics) for a range of network topologies. Mateo et al. reported a nontrivial relationship between the frequency of the driving signal and the optimal network topology~\cite{mateo2019}. The emergent collective response of the networked dynamical system to a slow-changing perturbation increases with the degree of the interaction network, while the opposite is true for the response to a fast-changing one. 

Although it is known that at high enough frequency the response is determined by the degree distribution~\cite{mateo2017}, a key question remains open: what is the importance of the structural properties---beyond just the degree distribution---of the network as predictors of a system's response at different frequencies. Mateo et al.~\cite{mateo2019} suggested to study the impact of other network characteristics, such as the clustering coefficient, but it is stated that it is not possible to change the degree distribution without also changing other characteristics, such as the clustering coefficient. For instance, Arenas et al.~\cite{arenas2008synchronization} stressed the existence of significant discrepancies in results in the literature for different network models when considering synchronization problems in complex networks. These reported discrepancies are attributed to the fact that multiple non-independent parameters (e.g. average degree, clustering coefficient, average shortest path) characterizing the network were concomitantly changed. This important issue stresses the challenges associated with any parametric studies of networked systems.

To avoid the pitfalls highlighted by Arenas et al. in Ref.~\cite{arenas2008synchronization} and to carry out a meaningful parametric study of networked systems with different graph topologies, one could look at pairs of networks with the same average degree, but with different clustering coefficients. Families of fully deterministic graphs, such as the generalized circulant graphs presented here, provide a useful tool in determining the impact of network structure on particular complex dynamical systems. Given their deterministic character, it can be known exactly how much a specific graph topological metric, such as the clustering coefficient, varies from one graph to the other, thereby making it easier to test and analyze new hypotheses. To this end, we now show how the classes of generalized circulant graphs introduced in the previous sections can be used to construct pairs of graphs of the same order (i.e. with the same number of nodes) and size (i.e. with the same number of edges). Such pairs of graphs thus have the same average degree, but as we will see different clustering coefficients. Let $G_1$ denote a generalized circulant graph $C_{N_1,k_1,m_1,1}$ of Type I and $G_2$ a generalized circulant graph $D_{N_2,k_2,m_2}$ of Type II. 
The graphs $G_1$ and $G_2$ have the same number of nodes, if the following condition is satisfied:
\begin{equation}\label{nodes}
N_1 + m_1 = N_2 + 1. 
\end{equation}
The condition for the two graphs to have the same number of links reads:
\begin{equation}\label{links}
\frac{N_1 k_1}{2}+m_1 = \frac{N_2 k_2}{2}+m_2.
\end{equation}
Solving Eq.~\eqref{nodes} for $N_2$, and substituting the solution into Eq.~\eqref{links} leads to
\begin{equation}\label{links2}
N_1(k_1-k_2)=(m_1-1)k_2 + 2m_2 - 2m_1.
\end{equation}
With the particular choice $k_1 - k_2 = 2$, the solution of the Diophantine equation~\eqref{links2} is given by
\begin{equation}\label{N1}
N_1=(m_1-1)\frac{k_2}{2} + m_2 - m_1.
\end{equation}
\begin{figure}[htbp]
\centering
\includegraphics[]{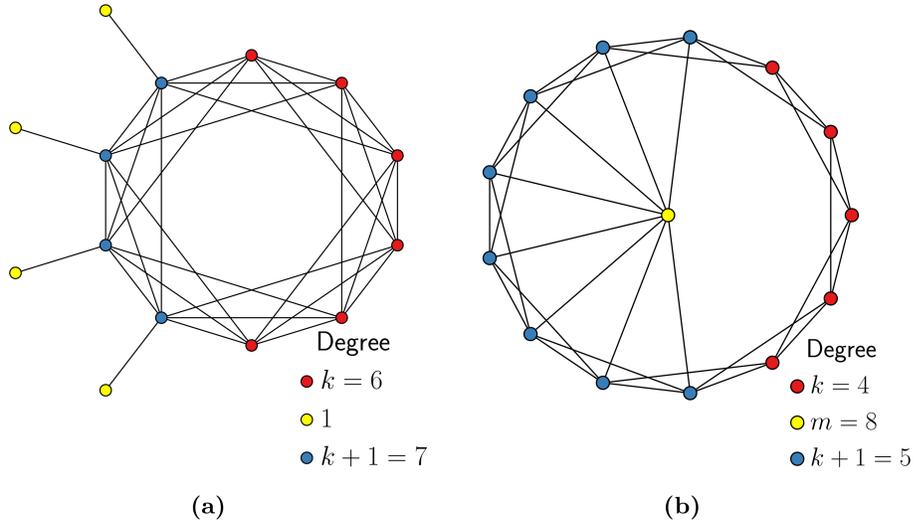}
\caption{Two generalized circulant graphs of the same order and size.Particular cases of $C_{10,6,4}$ (a) and $D_{13,4,8}$ (b).}\label{fig:12}   
\end{figure}
Note that $N_1$ is an integer because $k_2$ is even. Equation~\eqref{N1} gives rise to a family of pairs of graphs $C_{N_1,k_1,m_1,1}$ and $D_{N_2,k_2,m_2}$ of the same order and size. This family of graphs is characterized by three parameters. For example, the choice $m_1=4, m_2=8, k_2=4$, leads to the pair of graphs $C_{10,6,4}$ and $D_{13,4,8}$, both having 14 nodes and 34 links (see Fig.~\ref{fig:12}). However, these two graphs have notably different clustering coefficients: $C(C_{10,6,4})=0.380$ and $C(D_{13,4,8})=0.569$. It is worth noting that by altering the choice $k_1 - k_2 = 2$,  to $k_1 - k_2 = 2t$, with $t$ being an integer larger than 1, we can construct other families of graph pairs, of the same order and size as well.

\begin{figure}[htbp]
\centering  
\includegraphics[width=0.8\textwidth]{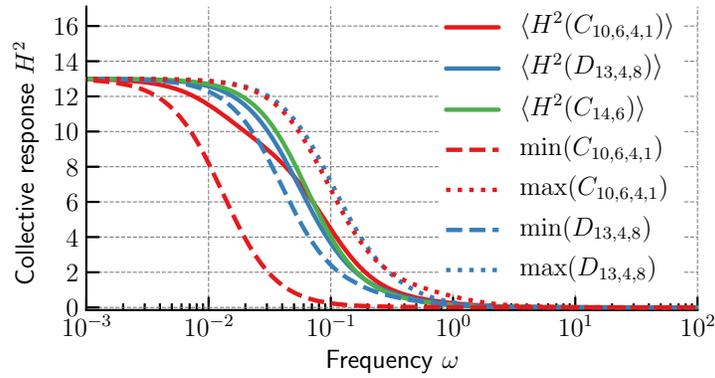}
\caption{Collective frequency response for graphs with similar number of edges and nodes. The graph $C_{14,6}$ with 42 edges is included for the sake of comparison.}\label{fig:13}
\end{figure}

We can then use the two graphs generated by this approach, to analyze the same distributed linear leader-follower consensus as in Ref.~\cite{mateo2019}. By using these graphs, we can study the impact of the clustering coefficient on the collective frequency response $H^2(\omega)$---metric that can be interpreted as the number of agents that are able to respond or follow a leader evolving at frequency $\omega$---but with equal average degree. Specifically, we follow the analytical approach outlined in Ref.~{\cite{mateo2019}} to compute the collective response $H^2$ (see Materials and Methods section in~\cite{mateo2019}). Since the graphs have the same number of nodes and edges,  the collective frequency responses for the two graphs can then readily be compared so as to quantify the impact of the clustering coefficient. The results for this pair of networks are shown in Fig.~{\ref{fig:13}}
Given that both graphs have 14 nodes, there are 14 different possible nodes that can be given the ``leader" role, hence we have a distribution of 14 collective frequency responses. In Fig.~\ref{fig:13}, we report the average collective response, $\langle H^2\rangle$, along with the minimum and maximum values. The average collective response for the classical circulant graph $C_{14,6}
$ is provided for reference. These results show that although $C_{10,6,4,1}$, $D_{13,4,8}$ and $C_{14,6}$ have the same average degree, they yield different average collective response $\langle H^2\rangle$. The differences are particularly marked in the mid-range of frequencies (i.e. for $5\times 10^{-2} <\omega< 10^{-1}$). It is quite clear that the low level of clustering of $C_{10,6,4,1}$ has a detrimental effect on the collective response of this networked system. Surprisingly though, the ``best" leader node---i.e. the node such that when driven yields the maximum collective frequency response---in both proposed graphs show similar levels of collective frequency response, for the best leader in both networks (dotted lines), but at different cost to the average agents response (solid lines).  

These conclusions for the collective response of these particular networked systems highlight the usefulness of the analytical results obtained for the clustering coefficients of the generalized circulant graphs of Type I, II and III. 

Even for small system sizes $N = O(10)$ where the statistical properties of grown networks are unreliable, the effect of the network topology has nonetheless a noticeable impact as appears clearly in Fig~{\ref{fig:13}}. This increases the usefulness of such deterministic network models, where network parameters can be varied greatly and known exactly, even for small networks. It is worth adding that many experiments with artificial networked systems are often on a limited scale, i.e. with a number of agents of the order of several tens~{\cite{mateo2019,kit2019,zoss2018}}. It is also possible to imagine that the new families of generalized circulant graphs introduced here could be used to study dynamical processes on more complex networks, such as multi-layered social networks~{\cite{manivannan2018different}} for instance.

\section*{Acknowledgments}
This work is partially supported by the National Research Foundation (NRF), Prime Minister's Office, Singapore, under its National Cybersecurity R\&D Programme (Award No. NRF2014NCR-NCR001-040), administered by the National Cybersecurity R\&D Directorate. The authors also acknowledge the support from the SUTD-MIT International Design Center (http://idc.sutd.edu.sg) under the Grant \#IDG31900101.

\section*{Authors Contribution}
R.K. and R.B. designed the study. R.K. and N.H. developed the analytical and numerical tools. All authors wrote the manuscript. 

\section*{Competing Interests}
The authors declare that they have no competing interests. 

\section*{Data and materials availability}
All data needed to evaluate the conclusions in the paper are present in the paper. Additional data related to this paper may be requested from the authors.

\section*{References}


\begin{thebibliography}{10}
\expandafter\ifx\csname url\endcsname\relax
  \def\url#1{\texttt{#1}}\fi
\expandafter\ifx\csname urlprefix\endcsname\relax\def\urlprefix{URL }\fi
\expandafter\ifx\csname href\endcsname\relax
  \def\href#1#2{#2} \def\path#1{#1}\fi

\bibitem{barrat2008dynamical}
A.~Barrat, M.~Barthelemy, A.~Vespignani, Dynamical processes on complex
  networks, Cambridge {U}niversity {P}ress, 2008.

\bibitem{watts1998collective}
D.~J. Watts, S.~H. Strogatz, Collective dynamics of `small-world' networks,
  Nature 393~(6684) (1998) 440.

\bibitem{holme2002growing}
P.~Holme, B.~J. Kim, Growing scale-free networks with tunable clustering, Phys.
  Rev. E 65~(2) (2002) 026107.

\bibitem{sekunda2016}
A.~Sekunda, M.~Komareji, R.~Bouffanais, Interplay between signaling network
  design and swarm dynamics, {N}etwork {S}cience 4~(2) (2016) 244--265.
\newblock \href {https://doi.org/10.1017/nws.2016.5}
  {\path{doi:10.1017/nws.2016.5}}.

\bibitem{kalna2007clustering}
G.~Kalna, D.~J. Higham, A clustering coefficient for weighted networks, with
  application to gene expression data, {AI} Communications 20~(4) (2007)
  263--271.

\bibitem{wang2017identifying}
Z.-Y. Wang, J.-T. Han, J.~Zhao, Identifying node spreading influence for
  tunable clustering coefficient networks, Physica A 486 (2017) 242--250.

\bibitem{centola2007cascade}
D.~Centola, V.~M. Egu{\'\i}luz, M.~W. Macy, Cascade dynamics of complex
  propagation, Physica A 374~(1) (2007) 449--456.

\bibitem{newman2001clustering}
M.~E. Newman, Clustering and preferential attachment in growing networks, Phys.
  Rev. E 64~(2) (2001) 025102.

\bibitem{MU201955}
J.~Mu, W.~Zheng, J.~Wang, J.~Liang,
  \href{http://www.sciencedirect.com/science/article/pii/S0950705119301698}{A
  novel edge rewiring strategy for tuning structural properties in networks},
  Knowledge-Based Systems 177 (2019) 55 -- 67.
\newblock \href {https://doi.org/https://doi.org/10.1016/j.knosys.2019.04.004}
  {\path{doi:https://doi.org/10.1016/j.knosys.2019.04.004}}.
\newline\urlprefix\url{http://www.sciencedirect.com/science/article/pii/S0950705119301698}

\bibitem{Kr_ger_2015}
B.~Krüger, E.~M. Schmidt, K.~Mecke,
  \href{https://doi.org/10.1088/1367-2630/17/2/023013}{Unimodular lattice
  triangulations as small-world and scale-free random graphs}, New Journal of
  Physics 17~(2) (2015) 023013.
\newblock \href {https://doi.org/10.1088/1367-2630/17/2/023013}
  {\path{doi:10.1088/1367-2630/17/2/023013}}.
\newline\urlprefix\url{https://doi.org/10.1088/1367-2630/17/2/023013}

\bibitem{articleShang}
Y.~Shang, Distinct clusterings and characteristic path lengths in dynamic
  small-world networks with identical limit degree distribution, Journal of
  Statistical Physics 149 (11 2012).
\newblock \href {https://doi.org/10.1007/s10955-012-0605-8}
  {\path{doi:10.1007/s10955-012-0605-8}}.

\bibitem{yang2014}
C.-X. Yang, M.-X. Tang, H.-Q. Tang, Q.-Q. Deng, Local-world and cluster-growing
  weighted networks with controllable clustering, International Journal of
  Modern Physics C 25 (11 2014).
\newblock \href {https://doi.org/10.1142/S0129183114400099}
  {\path{doi:10.1142/S0129183114400099}}.

\bibitem{Serrano}
M.~\'Angeles~Serrano, M.~Bogu\~n\'a,
  \href{https://link.aps.org/doi/10.1103/PhysRevE.72.036133}{Tuning clustering
  in random networks with arbitrary degree distributions}, Phys. Rev. E 72
  (2005) 036133.
\newblock \href {https://doi.org/10.1103/PhysRevE.72.036133}
  {\path{doi:10.1103/PhysRevE.72.036133}}.
\newline\urlprefix\url{https://link.aps.org/doi/10.1103/PhysRevE.72.036133}

\bibitem{BTER}
C.~Seshadhri, T.~G. Kolda, A.~Pinar,
  \href{https://link.aps.org/doi/10.1103/PhysRevE.85.056109}{Community
  structure and scale-free collections of {E}rd{\H{o}}s--{R}\'enyi graphs},
  Phys. Rev. E 85 (2012) 056109.
\newblock \href {https://doi.org/10.1103/PhysRevE.85.056109}
  {\path{doi:10.1103/PhysRevE.85.056109}}.
\newline\urlprefix\url{https://link.aps.org/doi/10.1103/PhysRevE.85.056109}

\bibitem{mateo2019}
D.~Mateo, N.~Horsevad, V.~Hassani, M.~Chamanbaz, R.~Bouffanais, Optimal network
  topology for responsive collective behavior, {S}cience {A}dvances 5 (2019)
  eaau0999.
\newblock \href {https://doi.org/10.1126/sciadv.aau0999}
  {\path{doi:10.1126/sciadv.aau0999}}.

\bibitem{zoss2018}
B.~M. Zoss, D.~Mateo, Y.~K. Kuan, G.~Toki{\'c}, M.~Chamanbaz, L.~Goh,
  F.~Vallegra, R.~Bouffanais, D.~K. Yue, Distributed system of autonomous buoys
  for scalable deployment and monitoring of large waterbodies, Autonomous
  Robots 42~(8) (2018) 1669--1689.

\bibitem{kit2019}
J.~L. Kit, A.~G. Dharmawan, D.~Mateo, S.~Foong, G.~S. Soh, R.~Bouffanais, K.~L.
  Wood, Decentralized multi-floor exploration by a swarm of miniature robots
  teaming with wall-climbing units, in: 2019 International Symposium on
  Multi-Robot and Multi-Agent Systems (MRS), IEEE, 2019, pp. 195--201.

\bibitem{newman2003structure}
M.~E. Newman, The structure and function of complex networks, SIAM Review
  45~(2) (2003) 167--256.

\bibitem{barrat2000properties}
A.~Barrat, M.~Weigt, On the properties of small-world network models, Eur.
  Phys. J. B 13~(3) (2000) 547--560.

\bibitem{circulants}
S.-M. Tang, Y.-L. Wang, C.-Y. Li, Generalized recursive circulant graphs,
  Parallel and Distributed Systems, IEEE Transactions on 23 (2012) 87 -- 93.
\newblock \href {https://doi.org/10.1109/TPDS.2011.109}
  {\path{doi:10.1109/TPDS.2011.109}}.

\bibitem{mateo2017}
D.~Mateo, Y.~K. Kuan, R.~Bouffanais, Effect of correlations in swarms on
  collective response, {S}cientific {R}eports 7~(1) (2017) 10388.
\newblock \href {https://doi.org/10.1038/s41598-017-09830-w}
  {\path{doi:10.1038/s41598-017-09830-w}}.

\bibitem{arenas2008synchronization}
A.~Arenas, A.~D{\'\i}az-Guilera, J.~Kurths, Y.~Moreno, C.~Zhou, Synchronization
  in complex networks, Physics Reports 469~(3) (2008) 93--153.

\bibitem{manivannan2018different}
A.~Manivannan, W.~Q. Yow, R.~Bouffanais, A.~Barrat, Are the different layers of
  a social network conveying the same information?, EPJ Data Science 7~(1)
  (2018) 34.

\end{thebibliography}

\end{document}